\documentclass[english,twoside,a4paper,10pt]{article} 

\usepackage[latin1]{inputenc}
\usepackage[T1]{fontenc}
\usepackage{amsmath}
\usepackage{amsfonts}
\usepackage{graphicx}
\usepackage{subfigure}
\usepackage{wrapfig}
\usepackage{a4wide}
\usepackage{amssymb}
\usepackage{fancyhdr}
\usepackage{mathrsfs}
\usepackage[toc,page]{appendix}

\def\ba{\begin{eqnarray}}
\def\ea{\end{eqnarray}}
\def\be{\begin{equation}}
\def\ee{\end{equation}}
\def\p{\partial}

\begin{document}

\title{\bf Black Hole entropy for two higher derivative theories of gravity}
\author{ Emilio~Bellini$^{1}$, Roberto~Di~Criscienzo$^{2}$\footnote{E-mail address:rdicris@science.unitn.it
}, Lorenzo~Sebastiani$^{2}$\footnote{E-mail address:l.sebastiani@science.unitn.it
} and
  Sergio~Zerbini$^{2}$\footnote{E-mail address:zerbini@science.unitn.it
}\\
\\
\begin{small}
$^{1}$ Dipartimento di Fisica ``G. Galilei'', Universit\`a di Padova\end{small}\\
\begin{small}and Istituto Nazionale di Fisica Nucleare - Sezione di Padova, Via Marzolo 8 -- 35131 Padova, Italia\end{small}\\
\begin{small}$^{2}$ Dipartimento di Fisica, Universit\`a di Trento\end{small}\\
\begin{small}and Istituto Nazionale di Fisica Nucleare - Gruppo Collegato di Trento,\end{small}\\
\begin{small}Via Sommarive 14 -- 38123 Povo, Italia\end{small}\\
}
\date{}

\maketitle

\def\thesection{\Roman{section}}
\def\theequation{\Roman{section}.\arabic{equation}}

\begin{abstract}{The dark energy issue is attracting the attention of
an increasing number of physicists all over the world. Among
the possible alternatives to explain what as been named the
\textquotedblleft Mystery of the Millennium'' are the so-called
Modified Theories of Gravity. A crucial test for such models is
represented by the existence and (if this is the case) the
properties of their black hole solutions. Nowadays, to our
knowledge, only two non-trivial, static, spherically symmetric,
solutions with vanishing cosmological constant are known by
Barrow \& Clifton (2005) and Deser, Sarioglu \& Tekin (2008).
The aim of the paper is to discuss some features of such
solutions, with emphasis on their thermodynamic properties such
as entropy and temperature.}
\end{abstract}




\section{Introduction}

Since the discovery by Riess and Perlmutter and respective
collaborators \cite{Riess:1998cb,perlmutter} that the universe
is---against any previous belief---in an accelerating epoch,
the dark energy issue has become the \textquotedblleft Mystery
of the Millennium'' \cite{Padmanabhan:2006ag}. Today, dark
energy is probably the most ambitious and tantalizing field of
research because of its implications in fundamental physics.
That the dark energy fluid has an equation of state index $w$
very close to minus one represents an important point in favour
of those who propose to explain dark energy in terms of a
cosmological constant, $\Lambda$. Still, a non-vanishing
cosmological constant does not exhaust the range of models that
have been proposed so far in order to solve the aforementioned
issue. This is justified, in part, by the whole sort of
well-known problems raised by the existence of a strictly
positive cosmological constant.

On the other hand, it is well accepted the idea according to
which general relativity is not the ultimate theory of gravity,
but an extremely good approximation valid in the present day
range of detection. It basically comes from this viewpoint the
input to so-called modified theories of gravity which nowadays
enjoy great popularity (\textit{cf}.
\cite{review1,review2,review3,review4,review5,review6,review7}
for a review). Without any claim for unification, such models
propose to change the Einstein--Hilbert Lagrangian to a more
general form able to reproduce the same general relativity
tests on solar distance scales and further justify both
inflationary and current acceleration of the universe.

The original idea of introducing a correction to the
Einstein--Hilbert action in the form of $f(R)=R+R^2$ was
proposed long time ago by Starobinsky \cite{Starobinsky:1980te}
in order to solve many of the problems left open by the
so-called hot universe scenario. This, in turn, had the
consequence of introducing an accelerating expansion in the
primordial universe, so that the Starobinsky model can be
considered as the first inflationary models. The recent
interest in models of modified gravity instead, grew up in
cosmology with the appearance of  \cite{turner,capo1,capo2}.

The mathematical structure of $f(R)$-theories of gravity
and their physical properties (e.g., asymptotic flatness,
renormalizability, unitarity) have been an exciting field of
research over the last four decades; a small but significant
trace of which is represented by
\cite{Utiyama,Fradkin1,Fradkin2,Stelle,Avramidi:1985ki,Avramidi2,Hawking:2001yt}.

The arena of models is in principle infinite while departures
from Einstein's theory are most of the times all but minimal.
Of crucial interest is, of course, the existence and, if this
is the case, the properties of black holes in modified
gravities. It is quite easy to find the conditions allowing
the existence of de Sitter-Schwarzschild black holes (see, for
example \cite{cogno05} for $f(R)$ modified gravity,
\cite{cogno06} for Gauss-Bonnet modified gravity, and
\cite{chile1,chile2,china,ita} for related topics).

Here, we are interested in non-trivial and static black holes
solutions. However, the number of exact non-trivial static
black hole solutions so far known in modified theories of
gravity is extremely small: just two, both spherically
symmetric. They have been obtained by Barrow \& Clifton (2005)
in a modified theory of the type $f(R)= R^{1+\delta}$ with
$\delta$ a small real parameter; and by Deser, Sarioglu \&
Tekin (2008) by adding to Einstein--Hilbert Lagrangian a
non-polynomial contribution of the type $\sqrt{C^2}$, with
$C_{a b c d}$ being the Weyl tensor.

These black hole solutions are not expected to share the same
laws of their Einsteinian counterparts: for this reason,
following \cite{Visser:1993nu}, we shall refer to them as
\textit{dirty black holes}. Some of the physical quantities one
would like to address to dirty black holes are their mass, the
horizon entropy, their temperature and so on. Thanks to the
large amount of work carried over in the last decade, we can
firmly say that the issue of entropy and temperature of dirty
black holes represents a well posed problem \cite{Wald:1993nt};
a nice and recent review on the entropy issue associated with
$f(R)$ gravity models is \cite{fara10}, where a complete list
of references can be found. Here, we only mention
\cite{vanzo,cogno05,cogno06}.  However, with regard to the
mass issue, all considerations still lay on a much more precise
ground.

In the present paper we shall work in units of
$c=G=\hbar=k_B=1$. The organization is as follows: in
\textsection 2 we review the Deser-Sarioglu-Tekin solution and
compute entropy and temperature for such black hole; in
\textsection 3 we do the same for the Clifton-Barrow solution.
In the Conclusions we address the difficulties faced trying to
define meaningfully the concept of mass for dirty black holes.

%

\section{The Deser-Sarioglu-Tekin solution}

Let us start by recapitulating the Deser-Sarioglu-Tekin
solution \cite{Deser:2007za}. The authors start from the action
\be I_{DST} = \frac{1}{16\pi} \int_{\mathcal M} d^4 x \sqrt{-g}
\left(R + \sqrt{3}\sigma \sqrt{C^2}\right) + \mbox{Boundary
Term} \label{d action} \ee where $\sigma$ is a real parameter
and $C^2 := C_{a b}^{\;\;\;c d} C_{c d}^{\;\;\;a b}$ is the
trace of the Weyl tensor squared. Looking for static,
spherically symmetric solutions of the type, \be ds^2 =
-a(r)b(r)^2 dt^2 + \frac{dr^2}{a(r)} + r^2 (d\theta^2 +
\sin^2\theta d\phi^2) \label{d metric} \ee the action (\ref{d
action}) becomes \be I_{DST}[a(r),b(r)] = \frac{1}{2} \int dt
\int_0^\infty dr \left[(1-\sigma)(r a(r) b'(r) + b(r)) + 3
\sigma a(r)b(r)\right]. \ee Imposing the stationarity condition
$\delta I[a(r),b(r)]=0$ gives the equations of motion for the
unknown functions $a(r)$ and $b(r)$. \ba
(1-\sigma)r b'(r) + 3 \sigma b(r) &=& 0 \nonumber \\
(1-\sigma) r a'(r) + (1-4\sigma)a(r) &=& 1-\sigma \label{d eom}
\;. \ea According to $\sigma$, the space of solutions of
(\ref{d eom}) can be different, in particular:
\begin{itemize}
\item $\sigma=0$ corresponds to Einstein-Hilbert action. In fact,
$a(r)= 1-\frac{\hat c}{r}$ and $b(r) = \hat k$ and for $\hat c,
\hat k$ positive constants, the Schwarzschild solution of
general relativity is recovered;

\item $\sigma =1$: only the trivial, physically unacceptable, solution $a(r)=0=b(r)$
exists;

\item $\sigma=\frac{1}{4}$: then, for some positive constants $\tilde k$ and $r_0$:
\be
a(r) = \ln\left(\frac{r_0}{r}\right) \qquad \mbox{and} \qquad b(r)=\frac{\tilde k}{r}\;;\label{d sol 2}
\ee

\item In all other cases, the general solution to (\ref{d eom}) turns out to be
\be a(r)= \frac{1-\sigma}{1-4\sigma} - c
r^{-\frac{1-4\sigma}{1-\sigma}} \qquad \mbox{and}\quad b(r)=
\left(\frac{r}{k}\right)^{\frac{3\sigma}{\sigma-1}}\label{d
sol}\;, \ee for some positive constants $c, k$.
\end{itemize}
The constants $\hat k, \tilde k$ and $k$ appearing in $b(r)$
are removable by time re-scaling. Notice also that, in (\ref{d
sol 2}), $g_{00}$ and $g_{11}$ go to zero as $r \rightarrow
\infty$ so that the model is unphysical. For this reason, we
shall mainly concentrate on the solution (\ref{d sol})
parametrized by all the $\sigma \neq 0, 1, \frac{1}{4}$. \\ In
order to treat (\ref{d sol}), let us introduce the parameter
$p(\sigma):= \frac{1-\sigma}{1-4\sigma}$ so that the metric
becomes \be ds^2 = -(p- c r^{-\frac{1}{p}})
\left(\frac{r}{k}\right)^{2\left(\frac{1-p}{p}\right)}dt^2 +
\frac{dr^2}{(p- c r^{-\frac{1}{p}})} + r^2 (d\theta^2 +
\sin^2\theta d\phi^2)\;.\label{d solution} \ee For $p<0$, or
$\frac{1}{4} < \sigma < 1$, $a(r)= - (\vert p \vert + c
r^{\frac{1}{\vert p\vert}}) < 0$ for all $r$, that is, the
parameter region $\frac{1}{4} < \sigma < 1$ needs to be
excluded to preserve the metric signature. As regard the
asymptotic behaviour of (\ref{d solution}), we see that:
\begin{itemize}
\item for $p>1$ or $0< \sigma <\frac{1}{4}$, we have that
$g_{00} \rightarrow 0$ and $g_{11}\rightarrow \frac{1}{p}$ as $r\rightarrow
\infty$;

\item for $0 < p <1$ or $\sigma \in (-\infty,0) \cup (1,+\infty)$,
we have that $g_{00} \rightarrow \infty$ and $g_{11}\rightarrow \frac{1}{p}$ as $r\rightarrow \infty$.
\end{itemize}
As noted by Deser \textit{et al}. the fact that the asymptotics
of $g_{00}$ and $g_{11}$ differ means that the equivalence
principle is violated: something which is intimately related
with the difficulty of defining a \textquotedblleft mass'' in
this theory \cite{Deser:2007za}.

Looking at the solution (\ref{d solution}), we see that the
hypersurface $r = r_H := \left(\frac{c}{p}\right)^p$ defined by
the condition $a(r_H)=0$ behaves as a Killing horizon with
respect to the timelike Killing vector field $\xi^a$. To prove
this, let us define a complex null tetrad $\{l^a, n^a, m^a,
\bar m^a\}$ for the metric (\ref{d solution}) according to the
following \linebreak rules  \cite{FN}:
\begin{enumerate}
\item $l^a$ is s.t. on the horizon
\be
l^a_H \equiv \xi^a\; ;
\ee
\item The normalization conditions hold
\be
l\cdot n =-1\ \qquad \&\qquad m\cdot \bar m=1\;;
\ee

\item All the other scalar products vanishes.
\end{enumerate}
Since the metric (\ref{d solution}) is not asymptotically flat,
it is not clear at all what is the right normalization for
$\xi^a$. Assuming $\xi^a=\lambda\, \p_t^{\;a}$, $\lambda \in
\mathbb R^+$, it's not difficult to see that \ba
l^a &=&  \left(\lambda,\lambda\, a(r)b(r),0,0\right)\;, \nonumber \\
n^a &=& \left(\frac{1}{2\lambda \,a(r) b(r)^2},-\frac{1}{2\lambda \,b(r)},0,0\right)\;,\nonumber \\
m^a &=& \left(0,0, \frac{i}{\sqrt{2} r}, \frac{1}{\sqrt{2} r\sin\theta}\right)\;, \nonumber \\
\bar m^a &=& \left(0,0, -\frac{i}{\sqrt{2} r },
\frac{1}{\sqrt{2} r\sin\theta}\right)\;, \label{cnt} \ea
satisfy the list of conditions to form a complex null tetrad.
As a consequence, for example, the metric can be re-written as
$g_{a b}= - 2 l_{(a} n_{b)} + 2 m_{(a} \bar m_{b)}\;$. The null
expansions are, by definition, \ba
\Theta_-:=\nabla_a n^a + n^a l_b \nabla_a n^b + n_b l^a \nabla_a n^b = - \frac{1}{\lambda\, r b(r)}\;, \nonumber\\
\Theta_+ := \nabla_a l^a + l^a n_b \nabla_a l^b + l_b n^a
\nabla_a l^b = \frac{2\lambda}{r} \,a(r) b(r)\;.\label{null
exp} \ea Thus, in-going light rays always converge ($\Theta_-
<0$ for all $r>0$); out-going light rays, instead, focus inside
the horizon ($\Theta_+ <0$ as $r<r_H$), diverge outside it
($\Theta_+ >0$ as $r>r_H$) and run in parallel at the horizon
($\Theta_+ \vert_H =0$). When they are slightly perturbed in
the in-direction (that is, along $n$), the out-going null ray
is absorbed inside the horizon $r_H$ as it is confirmed by the
fact that the in-going Lie derivative $\mathcal{L}_n \Theta_+
\vert_H= -\frac{1}{r_H^2} < 0 \;$ is everywhere negative.
Computing the convergence ($\varrho:= - m^a \bar m^b \nabla_b
l_{a}$) and the shear ($\varsigma := - m^a m^b \nabla_b l_a$)
of the null congruences at the horizon we can immediately check
they vanish, as expected for any Killing horizon. The Killing
surface gravity \be \kappa_H:= -l^a n^b \nabla_a l_b \vert_H =
\lambda a'(r)_H b(r)_H \label{Kill sg}\;, \ee turns out to
depend by the normalization of the Killing vector $\xi^a$. In
order to fix $\lambda$, we may implement the conical
singularity method. To this aim, let us start by the Euclidean
metric \be ds_E^2= + \frac{dr^2}{W(r)} + V(r) d\tau^2 +
r^2d\Omega^2 \,, \ee where we suppose that both $V(r)$ and
$W(r)$ have a structure like \be V(r) = (r-\tilde r) v(r)
\qquad \&\qquad W(r) = (r-\tilde r) w(r) , \ee with $v(r),
w(r)$ regular for $r>\tilde r$. $\tilde r$ may be identified
with some type of horizon close to which we are interested in
the behaviour of the metric. \be r-\tilde r \equiv \zeta x^2
\,, \ee with $\zeta$ a constant we are going to fix very soon.
\ba ds_E^2 &=& \frac{1}{w(r)}\left[\frac{dr^2}{r-\tilde r} + (r
- \tilde
  r) v(r) w(r) d\tau^2\right] + (\tilde r + \zeta x^2)^2d\Omega^2 \nonumber\\
&\stackrel{x\ll 1}{\approx}& \left(\frac{4\zeta}{w(\tilde r)} dx^2 +
\zeta v(\tilde r) x^2 d\tau^2 \right)  + \tilde r^2 d\Omega^2.
\ea
Let us choose $\zeta= w(\tilde r)/4$, the Euclidean metric takes the form
\be
ds_E^2 \approx dx^2 + x^2 d\left(\frac{\sqrt{v(\tilde r) w(\tilde
    r)}}{2}\tau \right)^2  + {\tilde r}^2 d\Omega^2 \;, \qquad x\ll
1.\label{con_sing1}
\ee
(\ref{con_sing1}) shows that close to the horizon ($r \approx \tilde
r$ or $x \ll 1$) the metric factorizes into $\mathcal{K}_2 \times
\mathbb{S}^{2}_{\;\tilde r}$ : $\mathcal{K}_2$ being the metric of flat
two-dimensional metric on behalf of identifying $x$ with the polar
distance and $\tau$ with the angular coordinate. However,
$\mathcal{K}_2$ is regular if and only if
\be
 \frac{\sqrt{v(\tilde r) w(\tilde r)}}{2}\tau \sim
 \frac{\sqrt{v(\tilde r) w(\tilde r)}}{2}\tau + 2\pi
\ee
or, in other words,
\be
\tau \sim \tau + \frac{4\pi}{\sqrt{v(\tilde r) w(\tilde r)}} \equiv
\tau + \beta \label{conic_temp} .
\ee
$\beta$ representing the (unique) $\tau$-period which allows to impose a
smooth flat metric on $\mathbb R^2$. \\

In Quantum Field Theory, the KMS propagator exhibits a
periodicity in time when the system is at finite temperature.
The period of the compactified time, $\beta$, is directly
related to the temperature of the system in Lorentzian
signature, through ($k_B=1$) \be T = \frac{1}{\beta} \,. \ee If
we assume the standard Hawking temperature formula, $T=
\kappa_H/2\pi$, the period $\beta$ in (\ref{conic_temp}) can be
re-written according to \be \kappa_H = \frac{\sqrt{V'(\tilde r)
W'(\tilde r)}}{2},\label{KSG} \ee which for the metric (\ref{d
metric}) reads $\kappa_H = \frac{1}{2} a'(r)_H b(r)_H$.
Comparison between the latter and (\ref{Kill sg}) fixes the
normalization of the Killing vector $\xi^a$ to be $\lambda =
\frac{1}{2}$. What is most important to us is that $\kappa_H
\neq 0$, so that we may conclude that the Killing horizon is of
the bifurcate type. We may anticipate that this is not the
unique surface gravity which can defined for a generic
spherically symmetric static black hole. For the sake of
simplicity, we shall postpone this discussion to the Conclusion
an alternative definition.

Given these preliminary remarks, we are now in the position to
apply Wald's argument \cite{Wald:1993nt} to derive the black
hole entropy associated to the Killing horizon of the solution
(\ref{d solution}).\\ Following
\cite{Wald:1993nt,Visser:1993nu,Brustein:2007jj}, the explicit
calculation of the black hole entropy $S_W$ of the horizon \linebreak
$r=r_H= (c/p)^p$ is provided by the formula \be S_W = - 2\pi
\oint_{\tiny{\begin{array}{cc}
                     r= & r_H \\
             t = & \mbox{const}
                    \end{array}}} \left(\frac{\delta \mathscr L}{\delta R_{a b c d}}\right)^{(0)}\, \hat \epsilon_{a b} \,\hat \epsilon_{c d} \sqrt{h_{(2)}}\; d\theta \,d\phi \;,\label{wald}
\ee where $\mathscr L = \mathscr L(R_{a b c d}, g_{a b},
\nabla_a R_{b c d e},\dots)$ is the Lagrangian density of any
general theory of gravity, in the specific case, \be \mathscr
L(R_{a b c d}, g_{a b}, \nabla_a R_{b c d e},\dots) =
\frac{1}{16\pi}( R + \sqrt{3}\sigma \sqrt{C^2})\;. \ee The
hatted variable, $\hat \epsilon_{a b}$, is the binormal vector
to the (bifurcate) horizon: it is antisymmetyric under the
exchange of $a \leftrightarrow b$ and normalized so that $\hat
\epsilon _{a b} \hat \epsilon^{a b} =-2$. For the metric
(\ref{d metric}), the binormal turns out to be \be \hat
\epsilon_{a b} = b(r) (\delta^0_a \,\delta^1_b - \delta^1_a
\,\delta^0_b )\;. \ee The induced volume form on the bifurcate
surface $r=r_H$, $t=$constant is represented by
$\sqrt{h_{(2)}}\; d\theta \,d\phi$, where, for any spherically
symmetric metric, $\sqrt{h_{(2)}} = r^2 \sin\theta$  and the
angular variables $\theta, \phi$ run over the intervals
$[0,\pi]$, $[0, 2\pi)$, respectively.\\
Finally, the superscript $(0)$ indicates that the partial
derivative $\delta \mathscr L/\delta R_{a b c d}$ is evaluated
on shell. The variation of the Lagrangian density with respect
to $R_{a b c d}$ is performed as if $R_{a b c d}$ and the
metric $g_{a b}$ are independent.\\ In the specific case,
equation (\ref{wald}) becomes \be S_W = - 8\pi \mathscr A_H \,
b^2(r_H)\,\left(\frac{\delta \mathscr L}{\delta R_{0 1 0
1}}\right)^{(0)} \label{wald1}\;, \ee with $\mathscr A_H$ the
area of the black hole horizon. Let us compute the Lagrangian
variation, \ba
16\pi \,(\delta\mathscr L) &=& \delta R + \sqrt{3} \,\sigma \,\delta (\sqrt{C^2})\nonumber\\
&=& \frac{1}{2} (g^{a c} g^{b d} - g^{a d} g^{b c}) \delta R_{a b c d} + \frac{\sqrt{3} \sigma}{2} (C^2)^{-\frac{1}{2}}\, \delta (C^2)\;.
\ea
Using the fact that $C^2 = R_{a b c d} R^{a b c d} - 2 R_{a b} R^{a b} + \frac{1}{3} R^2$, we get,
\ba
\frac{\delta \mathscr L}{\delta R_{a b c d}}&=& \frac{1}{16\pi}\left\lbrace \frac{1}{2} (g^{a c} g^{b d} - g^{a d} g^{b c})  + \frac{\sqrt{3} \sigma}{2} (C^2)^{-\frac{1}{2}}\,\cdot \right.\nonumber \\
& & \left. \cdot \left[ 2 R^{a b c d} - (g^{a c} R^{b d} + g^{b
d} R^{a c} - g^{a d} R^{b c} - g^{b c} R^{a d}) + \frac{1}{3}
(g^{a c} g^{b d} - g^{a d} g^{b c }) R \right]\right\rbrace \;.
\label{variation} \ea In the specific, \be \left(\frac{\delta
\mathscr L}{\delta R_{0 1 0 1}}\right)^{(0)} = \frac{1}{32\pi}
\left[ g^{00} g^{11} + \frac{\sqrt{3}
\sigma}{\sqrt{C^2}}\left(2 R^{0 1 0 1} - g^{00} R^{1 1} - g^{1
1} R^{0 0} + \frac{1}{3} g^{0 0} g^{1 1}
R\right)\right]\Big\vert_H \,. \label{wald2} \ee Since in
general, tr $ C^n = \left(-\frac{1}{3}\right)^n [2 +
(-2)^{2-n}] X^n$, for $n>0$ and \be X(r) = \frac{1}{r^2} [r^2
a'' + 2(a-1) -2 r a'] + \frac{1}{r b} [3 r a' b' -2 a (b' - r
b'')] \ee for the metric (\ref{d metric}), we may write \be
\sqrt{C^2}\vert_H = \frac{1}{\sqrt{3}} \Big \vert \frac{1}{r^2}
[r^2 a'' + 2(a-1) -2 r a'] + \frac{1}{r b} [3 r a' b' -2 a (b'
- r b'')]\Big\vert_H \;.\label{wald3} \ee Taking together
(\ref{wald1}), (\ref{wald2}) and (\ref{wald3}), for both the
solutions (\ref{d sol 2}) and (\ref{d solution}), we finally
have that the horizon entropy for the Deser \textit{et al.}
black hole is \be S_W = \frac{\mathscr A_H}{4} \left(1 +
\varepsilon \sigma\right) \;,\qquad \mbox{where}\qquad
\varepsilon := \left\lbrace\begin{array}{cc}
             +1 ,& \quad \sigma \leq \frac{1}{4} \\
             -1,& \quad\sigma >1\;
         \end{array}\right.\;. \label{result}
\ee

\begin{figure}[h!]
\centering
  \includegraphics[width=13cm]{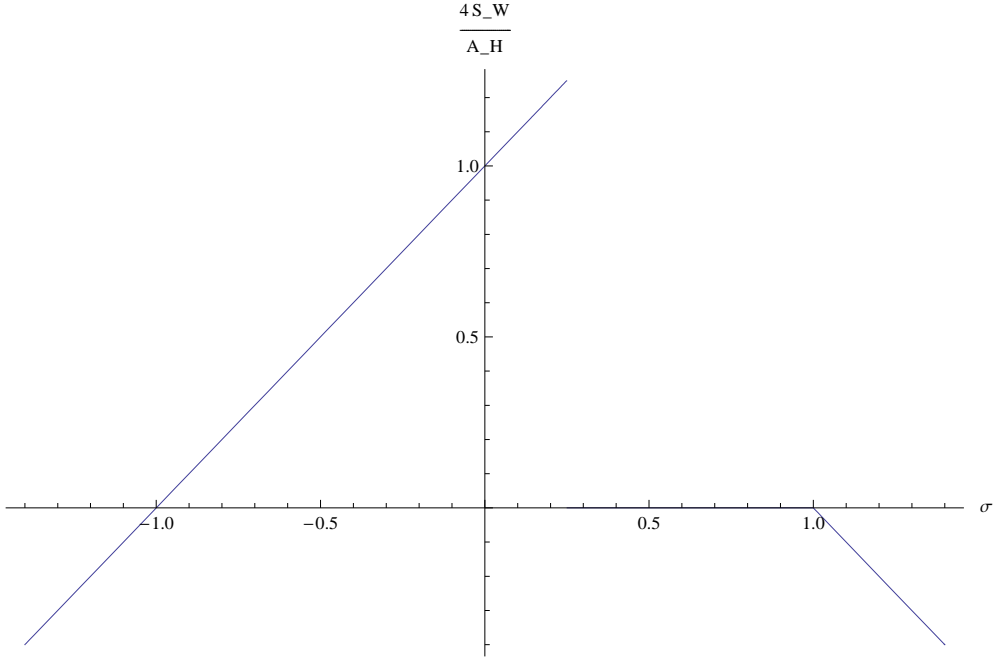}
\caption{\small{Wald's entropy in units of $\mathscr A_H/4$ versus $\sigma$ parameter for the Deser \textit{et al.} black hole.}}
\end{figure}
\pagebreak

According to (\ref{result}) the entropy predicted by Wald's
formula restricts considerably the space of the $\sigma$
parameter with respect to our previous considerations. In fact,
as shown by Figure 1, the entropy of the black hole is positive
only as far as $\sigma \in (-1, \frac{1}{4}]$. For $\sigma
=-1$, the entropy vanishes suggesting (but we leave this to the
level of a speculation) that, for this value of $\sigma$, the
number of microscopic configurations realizing the black hole
is only one. For $\sigma \in (-1,0)$, the entropy of Deser's
black hole is always smaller than its value in general
relativity. Notice also that for $\sigma=\frac{1}{4}$, the
entropy function is continuous even if the black hole metric
changes. However, as pointed out above, such solution is not
physical because of its pathological asymptotic behaviour.  \\
\textit{En passant}, we notice how Wald's entropy could be
computed equally well following \cite{Brustein:2007jj}.
Introducing a new radial co-ordinate $\rho$ such that \be
\rho(r) := \frac{k^{-\frac{1-p}{p}}}{p}\, r^{\frac{1}{p}} \ee
the metric (\ref{d solution}) transforms to \be ds^2=
-h(\rho)dt^2 + \frac{d\rho^2}{h(\rho)} + q(\rho) d\Omega^2
\label{brustein} \ee with \be h(\rho) = \left(\frac{p
\rho}{k}\right)^{2(1-p)} \left(p - \frac{c}{p k^{\frac{1-p}{p}}
\rho}\right)\;, \qquad q(\rho) = (p k^{\frac{1-p}{p}}
\rho)^{2p}. \ee This time, Wald's entropy (\ref{result}) will
follow from \be S_W = - 8\pi  \oint_{\tiny{\begin{array}{cc}
                     r= & r_H \\
             t = & \mbox{const}
                    \end{array}}} \left(\frac{\delta \mathscr L}{\delta R_{\rho t \rho t}}\right)^{(0)}\, q(\rho)\; d\Omega^2\;.\label{wald21}
\ee


\section{The Clifton-Barrow solution}

The Clifton-Barrow solution starts from the following
modified-gravity action (evaluated in the vacuum space):
\begin{equation}
I_{CB}=\int_{\mathcal{M}}d^4 x \sqrt{-g}\left(\frac{R^{1+\delta}}{\chi}\right)\,.
\end{equation}
Here, $\delta$ is a constant and $\chi$ is a dimensional
parameter. We can choose $\chi=16\pi G^{1+\delta}$. When
$\delta=0$, we recover the Hilbert-Einstein action of General
Relativity.

Taking the variation of the action with respect to the metric $g_{\mu\nu}$, we obtain:
\begin{equation}
R_{\mu\nu}=\delta\left(\frac{\partial^{\sigma}\partial^{\tau}R}{R}-(1-\delta)\frac{\partial^{\sigma}R\partial^{\tau}R}{R^2}\right)\left(g_{\mu\sigma}g_{\nu\tau}+\frac{1+2\delta}{2(1-\delta)}g_{\mu\nu}g_{\sigma\tau}\right)\,. \label{EqCampo}
\end{equation}
Looking for static, spherically symmetric metric of the type,
\begin{equation}
ds^2=-V(r)dt^2+\frac{dr^2}{W(r)}+r^2d\Omega^2\,,
\end{equation}
we find the Clifton-Barrow solution of Equation (\ref{EqCampo}):
\begin{equation}
V(r)=\left(\frac{r}{r_0}\right)^{2\delta(1+2\delta)/(1-\delta)}\left(1+\frac{C}{r^{(1-2\delta+4\delta^2)/(1-\delta)}}\right)\,,
\end{equation}
\begin{equation}
W(r)=\frac{(1-\delta)^2}{(1-2\delta+4\delta^2)(1-2\delta-2\delta^2)}\left(1+\frac{C}{r^{(1-2\delta+4\delta^2)/(1-\delta)}}\right)\,.
\end{equation}
$C$ and $r_{0}$ are dimensional constants.

In a similar way with respect to the previous section, we can
see that the hypersurface \linebreak
$r=r_{H}:=(-C)^{(1-\delta)/(1-2\delta+4\delta^2)}$, for which
$W(r_{H})=0$ and $\partial_{r}W(r_{H})\neq 0$, determines an
event horizon, and, since $C<0$, the Clifton-Barrow metric is a
Black Hole solution.

According to Equation (\ref{KSG}), we recover the Killing-horizon surface gravity
\begin{equation}
\kappa_H=\frac{1}{2}\sqrt{\frac{(1-2\delta+4\delta^2)}{(1-2\delta-2\delta^2)}}\frac{r_{H}^{(2\delta+2\delta^2-1)/(1-\delta)}}{r_0^{\delta(1+2\delta)/(1-\delta)}}\,,
\end{equation}
which can be used to find the Hawking temperature $T=\kappa_H/2\pi$.

As a last remark, we are able to derive the Black Hole entropy
associated to the event horizon of the Clifton-Barrow solution.
For modified gravity $F(R)$-theories (where the gravity
lagrangian is a function $F(R)$ of the Ricci scalar only), it
is easy to see that the Wald formula in Equation (\ref{wald}) is
simplified as
\begin{equation}
S_W=4\pi \mathcal{A}_H \frac{d F(R)}{d R}\Big\vert_{r_H}\,.
\end{equation}
In our case, $F(R)=R^{1+\delta}/\chi$, so we find:
\begin{equation}
S_{W}=\frac{4\pi\mathcal{A}_H}{\chi}(1+\delta)\left[\frac{6\delta(1+\delta)}{(2\delta^2+2\delta-1)r_H^2}\right]^\delta\,, \label{result2}
\end{equation}
proved by the fact that on the Clifton-Barrow solution
$R=6\delta(1+\delta)/((2\delta^2+2\delta-1)r^2)$.

In order to have the positive sign of entropy, we must require
$\delta>(\sqrt{3}-1)/2$. The solutions with
$0<\delta<(\sqrt{3}-1)/2$ are unphysical, whereas for
$\delta=0$ we find the result of General Relativity.

\section{Conclusions}

Despite the great success enjoyed by modified theories of
gravity, we have seen that only two non-trivial, static,
spherically symmetric, vacuum black-hole solutions are known so
far. Their thermodynamic properties have been taken into
considerations. We have shown that the solutions we considered
in \textsection 2 and 3 possess a Killing horizon with a
Killing vector $\xi^a \sim \p_t^a$ associated which cannot be
defined unambiguously due to the fact that the spacetimes are
not asymptotically flat. What is most important, however, is
that we have been able to deduce a non-vanishing Killing
temperature for such horizons. Of course, this is not the only
temperature we can define for such horizons. As shown in
\cite{Kodama}, in spherically symmetric spacetimes always
exists a Kodama vector field $K$ whose defining property is
that $(G_{a b}K^b)^{; a}=0$. The Kodama vector turns out to be
time/light/space-like in untrapped/marginal/trapped spacetime
regions; it gives a preferred flow of time generalizing the
Killing time flow familiar to static cases; it makes possible
to define an invariant particle energy even in non-stationary
spacetimes and it associates a Kodama-Hayward surface gravity
to any future outer trapping horizon \cite{sean}. In static,
asymptotically flat spacetimes, both the Killing and Kodama
vectors coincide, so that they give rise to the same concepts
of energy and temperature. In static, non-asymptotically flat
spaces, they are both ambiguous and can differ by
normalizations, but nonetheless the ratio \textquotedblleft
energy/surface gravity'' remains fixed \cite{noi1,noi2,noi3}.
This means that as far as the Killing temperature associated
with the black holes mentioned here is non-vanishing, also
their Kodama-Hayward temperature will be so. On the other hand,
that the horizons we are concerned are of the bifurcate type
means we are in Wald's hypothesis in order to compute their
entropy. In this sense, equations (\ref{result}--\ref{result2}) and Figure 1 represent our main
results.

To complete the picture of thermodynamic features of black
holes in modified theories of gravity, it would be necessary to
formulate a consistent definition of their mass. As it is well
known, in modified theories of gravity the first law of
thermodynamics generally requires a work term even in vacuum
solutions something which makes the first law useless in the
situations at hand. Quite recently some attempts have been put forward in order to answer
the question, but only for asymptotically flat spacetimes,
\textit{cf}. \cite{Deser:2007vs,av}.

\makebox[17.0cm][s]{In principle, a powerful tool to evaluate
the black hole mass in a theory of the type $\mathscr L= R +
(\dots)$} \newline is represented by the so-called Brown-York
quasi-local mass \cite{Chan1,Chan2,Chan3}. In static,
spherically symmetric spacetimes where the metric can be put in
the form (\ref{d metric}) the BY mass reads \be M_{BY} = r a(r)
b(r) \left[\sqrt{\frac{a^{(0)}(r)}{a(r)}} -1\right]\,
\label{by} \ee with $a^{(0)}(r)$ an arbitrary function which
determines the zero of the energy for a background spacetime
and $r$ is the radius of the spacelike hypersurface boundary.
When the spacetime is asymptotically flat, the ADM mass $M$ is
the $M_{BY}$ determined in (\ref{by}) in the limit $r
\rightarrow \infty$. If no cosmological horizon is present, the
large $r$ limit of  (\ref{by}) is used to determine the mass.
However, this approach is known to fail whenever the matter
action (\emph{i.e.} what we have represented with $(\dots)$ few line
above) contains derivatives of the metric as it is the case of
the Deser \textit{et al.} action, (\ref{d action}).

Another quasi-local energy definition well known in general
relativity and fully employed in spherical symmetry is the
so-called Misner-Sharp energy \cite{ms} which can be proved to
be the conserved charge generated by the Kodama vector $K$
\cite{Hayward:1993ph,Hayward:1996ph}. In the last few years,
different authors have tried to generalize the Misner-Sharp
energy definition to wider classes of gravity theories
\cite{Maeda:2007uu,cai}. But even if Cai \textit{et al.}
provide a general formula for the generalized MS energy in
$f(R)$ gravity, this does not produce any explicit, useful,
result for the Clifton-Barrow black hole.

In conclusion, we have succeeded in computing two of three most
relevant thermodynamic parameters (entropy and temperature) of
the known black hole solutions in modified theories of gravity;
the mass resisting up to now to any attack led by conventional
methods.

\section*{Acknowledgements}

E.B. thanks the members of the Group of Theory of Gravity in the
Department of Physics of the University of Trento where most of
this work has been done.


\bibliographystyle{mdpi}
\makeatletter
\renewcommand\@biblabel[1]{#1.}
\makeatother

\begin{thebibliography}{99}

\bibitem{Riess:1998cb}
  Riess, A.G.; Filippenko, A.V.; Challis, P.; Clocchiatti, A.; Diercks, A.; Garnavich, P.M.; Gilliland, R.L.; Hogan, C.J.; Jha, S.; Kirshner, R.P.; Leibundgut, B.;  Phillips, M.M.;  Reiss, D.; Schmidt, B.P.; Schommer, R.A.; Smith, R.S.; Spyromilio, J.; Stubbs, C.; Suntzeff N.b.; Tonry, J.
   Observational Evidence from Supernovae for an Accelerating Universe and a Cosmological Constant.
  {\em Astron.\ J.}\ {\bf 1998}, {\em 116}, 1009.

\bibitem{perlmutter}
 Perlmutter, S.;  Aldering, G.; Goldhaber, G.;  Knop, R.A.;  Nugent, P.;  Castro, P.G.;  Deustua, S.;  Fabbro, S.; Goobar, A.;  Groom, D.E.;  Hook, I.M.;  Kim, A.G.;  Kim, M.Y.;  Lee, J.C.; Nunes, N.J.;  Pain, R.;  Pennypacker, C.R.; Quimby,  R.; Lidman,  C.;  Ellis, R.S.;  Irwin, M.;  McMahon, R.G.;  Ruiz-Lapuente, P.;  Walton, N.;  Schaefer, B.; Boyle,  B.J.;  Filippenko, A.V.;  Matheson, T.;  Fruchter, A.S.; Panagia, N.;   Newberg, H.J.M.;  Couch, W.J.
 Measurements of $\Omega$ and $\Lambda$ from 42 high-redshift supernovae. {\em Astrophys.\ J.}\ {\bf 1999}, {\em 517}, 565 (1999).

\bibitem{Padmanabhan:2006ag}
  Padmanabhan, T. Dark energy: Mystery of the millennium. {\em AIP Conf.\ Proc.}\ {\bf 2006}, {\em 861}, 179.

\bibitem{review1}
 Sahni, V.; Starobinsky, A.A. The Case for a positive cosmological Lambda term. {\em Int. J. Mod. Phys.} {\bf 2000}, {\em D 9}, 373.

\bibitem{review2}
 Carroll, S.M.  The Cosmological constant. {\em Living Rev. Rel.} {\bf 2001}, {\em 4}, 1.

\bibitem{review3}
 Peebles, P.J.E.; Ratra, B. The cosmological constant and dark energy. {\em Rev. Mod. Phys.} {\bf 2003}, {\em 75}, 559.

\bibitem{review4}
 Padmanabhan, T. Cosmological constant-the weight of the vacuum. {\em Phys.\ Rept.}\ {\bf 2003}, {\em 380}, 235.

\bibitem{review5}
 Copeland, E.; Sami,M.; Tsujikawa, S. Dynamics of dark energy. {\em Int.\ J.\ Mod.\ Phys.}\ {\bf 2006}, {\em D15}, 1753.

\bibitem{review6}
 Nojiri, S.; Odintsov, S.D. Introduction to modified gravity and gravitational alternative for dark energy. {\em Int.\ J.\ Geom.\ Meth.\ Mod.\ Phys.}\ {\bf 2007}, {\em 4}, 115.

\bibitem{review7}
 Sotiriou, T.P.; Faraoni, V. $f(R)$ Theories Of gravity. {\em Rev. Mod. Phys.} {\bf 2010}, {\em 82}, 451.

\bibitem{Starobinsky:1980te}
  Starobinsky, A.A. A new type of isotropic cosmological models without singularity. {\em Phys.\ Lett.}\ {\bf 1980}, {\em B91}, 99.

\bibitem{turner}
 Carroll, S.M.; Duvvuri, V.; Trodden, M.; Turner, M.S. Is cosmic speed-up due to new gravitational physics? {\em Phys. Rev.} {\bf 2004}, {\em D70}, 043528.

\bibitem{capo1}
 Capozziello, S.; Carloni, S.; Troisi, A. Quintessence without scalar fields. {\em arXiv} astro-ph/0303041.

\bibitem{capo2}
 Capozziello, S.; Cardone, V.F.; Troisi, A. Reconciling dark energy models with $f(R)$ theories. {\em Phys. Rev.} {\bf 2005}, {\em D71}, 043503.

\bibitem{Utiyama}
  Utiyama, R.; DeWitt, B.S. Renormalization of a classical gravitational field interacting with quantized matter fields. {\em J.\ Math.\ Phys.} {\bf 1962}, {\em 3}, 608.

\bibitem{Fradkin1}
  Fradkin, E.S.; Vilkovisky, G.A. S Matrix for Gravitational Field. II. Local measure; General relations; Elements of renormalization theory. {\em Phys.\ Rev.}\ {\bf 1973}, {\em D8}, 4241.

\bibitem{Fradkin2}
Fradkin, E.S.; Tseytlin, A.A. Renormalizable asymptotically free quantum theory of gravity. {\em Nucl.\ Phys.}\ {\bf 1982}, {\em B201}, 469.

\bibitem{Stelle}
  Stelle, K.S. Renormalization of higher-derivative quantum gravity. {\em Phys.\ Rev.}\ {\bf 1977}, {\em D 16}, 953.

\bibitem{Avramidi:1985ki}
  Avramidi, I.G.; Barvinsky, A.O. Asymptotic freedom in higher derivative quantum gravity. {\em Phys.\ Lett.} {\bf 1985}, {\em B159}, 269.

\bibitem{Avramidi2}
 Avramidi, I.G. Asymptotic behavior of the quantum theory of gravity with higher order derivatives. {\em Yad.\ Fiz.}\ {\bf 1986}, {\em 44}, 255.

\bibitem{Hawking:2001yt}
  Hawking, S.W.; Hertog, T. Living with ghosts. {\em Phys.\ Rev.} {\bf 2002}, {\em D65}, 103515.

\bibitem{cogno05}
  Cognola, G.; Elizalde, E.; Nojiri, S.; Odintsov, S.D.; Zerbini, S. One-loop $f(R)$ gravity in de Sitter universe.  {\em JCAP} {\bf 2005}, {\em 0502}, 010.

\bibitem{cogno06}
  Cognola, G.; Elizalde, E.; Nojiri, S.; Odintsov, S.D.; Zerbini, S. Dark energy in modified Gauss-Bonnet gravity: Late-time acceleration  and the hierarchy problem. {\em Phys.\ Rev.} {\bf 2006}, {\em D73}, 084007.

\bibitem{chile1}
   Oliva, J.; Ray, S.A new cubic theory of gravity in five dimensions: Black hole, Birkhoff's theorem and C-function. {\em arXiv} 1003.4773.

\bibitem{chile2}
  Oliva, J.; Ray, S.A Classification of Six Derivative Lagrangians of Gravity and Static Spherically Symmetric Solutions. {\em arXiv} 1004.0737.

\bibitem{china}
 Cai, Y.F.; Easson, D.A. Black holes in an asymptotically safe gravity theory with higher derivatives. {\em JCAP} {\bf 2010}, {\em 1009}, 002.

\bibitem{ita}
 Berezhiani, Z.; Comelli, D.; Nesti, F.; Pilo, L. Exact spherically symmetric solutions in massive gravity. {\em JHEP} {\bf 2008}, {\em 0807}, 130.

\bibitem{Deser:2007za}
  Deser, S.; Sarioglu, O.; Tekin, B. Spherically symmetric solutions of Einstein + non-polynomial gravities. {\em Gen.\ Rel.\ Grav.} {\bf 2008}, {\em 40}, 1.

\bibitem{Clifton:2005aj}
  Clifton, T.; Barrow, J.D. The power of general relativity. {\em Phys.\ Rev.}\ {\bf 2005}, {\em D72}, 103005.

\bibitem{Wald:1993nt}
  Wald, R.M. Black hole entropy is the Noether charge. {\em Phys.\ Rev.} {\bf 1993}, {\em D48}, 3427.

\bibitem{Visser:1993nu}
  Visser, M. Dirty black holes: Entropy as a surface term. {\em Phys.\ Rev.}\ {\bf 1993}, {\em D48}, 5697.

\bibitem{fara10}
  Faraoni, V. Black hole entropy in scalar-tensor and $f(R)$ gravity: an overview. {\em Entropy} {\bf 2010}, {\em 12}, 1246.

\bibitem{vanzo}
  Brevik, I.H.; Nojiri, S.; Odintsov, S.D.; Vanzo, L. Entropy and universality of Cardy-Verlinde formula in dark energy
  universe. {\em Phys.\ Rev.} {\bf 2004}, {\em D70}, 043520.

\bibitem{Brustein:2007jj}
  Brustein, R.; Gorbonos, D.; Hadad, M. Wald's entropy is equal to a quarter of the horizon area in units of the effective gravitational coupling. {\em Phys.\ Rev.} {\bf 2009}, {\em D79}, 044025.

\bibitem{FN}
 Frolov, V.P.; Novikov, I.D. \textit{Black hole physics: Basic concepts and new developments}; Kluwer Academic: Dordrecht, Netherlands, 1998.

\bibitem{Kodama}
 Kodama, H. Conserved energy flux for the spherically symmetric system and the back reaction problem in the black hole evaporation. {\em Prog.\ Theor.\ Phys.} {\bf 1980}, {\em 63}, 1217.

\bibitem{sean}
 Hayward, S.A. Unified first law of black-hole dynamics and relativistic thermodynamics. {\em Class.\ Quant.\ Grav.}\ {\bf1998}, {\em 15}, 3147.

\bibitem{noi1}
  Di Criscienzo, R.; Nadalini, M.; Vanzo, L.; Zerbini, S.; Zoccatelli, G. On the Hawking radiation as tunnellingfor a class of dynamical black holes. {\em Phys.\ Lett.} {\bf 2007}, {\em B657}, 107.

\bibitem{noi2}
 Hayward, S.A.; Di Criscienzo, R.; Vanzo, L.; Nadalini, M.; Zerbini, S. Local Hawking temperature for dynamical black holes. {\em Class.\ Quant.\ Grav.} {bf 2009}, {\em 26 }, 062001.

\bibitem{noi3}
 Di Criscienzo, R.; Hayward, S.A.; Nadalini, M.; Vanzo, L.; Zerbini, S. Hamilton--Jacobi tunneling method for dynamical horizons in different coordinate gauges. {\em Class.\ Quant.\ Grav.}\ {\bf 2010}, {\em 27}, 015006.

\bibitem{Deser:2007vs}
   Deser, S.; Tekin, B. New energy definition for higher curvature gravities. {\em Phys.\ Rev.} {\bf 2007}, {\em D75}, 084032.

\bibitem{av}
 Abreu, G.; Visser, M. Tolman mass, generalized surface gravity, and entropy bounds. {\em Phys.\ Rev.\ Lett.}\ {\bf 2010},  {\em 105}, 041302.

\bibitem{Chan1}
 Brown, J.D.; York, J.W. Quasilocal energy and conserved charges derived from the gravitational action. {\em Phys. Rev.} {\bf 1993}, {\em D47}, 1407.

\bibitem{Chan2}
 Brown, J.D.; Creighton, J.; Mann, R.B.  Temperature, energy and heat capacity of asymptotically anti-de Sitter black holes. {\em Phys. Rev.} {\bf 1994}, {\em D50}, 6394.

\bibitem{Chan3}
 Chan, K.C.K.; Horne, J.H.; Mann, R.B. Charged dilaton black holes with unusual asymptotics. {\em Nucl.\ Phys.} {\bf 1995}, {\em B447}, 441.

\bibitem{ms}
 Misner, C.W.; Sharp, D.H.  Relativistic equations for adiabatic, spherically symmetric gravitational collapse. {\em Phys. Rev} {\bf 1964}, {\em 136}, B571.

\bibitem{Hayward:1993ph}
   Hayward, S.A. Quasilocal gravitational energy. {\em Phys.\ Rev.} {\bf 1994}, {\em D49}, 831.

\bibitem{Hayward:1996ph}
 Hayward, S. A. Gravitational energy in spherical symmetry. {\em Phys.\ Rev.} {\bf 1996}, {\em D53}, 1938.

\bibitem{Maeda:2007uu}
   Maeda, H.; Nozawa, M. Generalized Misner-Sharp quasi-local mass in Einstein-Gauss-Bonnet gravity.
  {\em Phys.\ Rev.}\ {\bf 2008}, {\em D77}, 064031.

\bibitem{cai}
 Cai, R.G.; Cao, L.M.; Hu, Y.P.; Ohta, N. Generalized Misner-Sharp Energy in $f(R)$ Gravity. {\em Phys.\ Rev.}\ {\bf 2009}, {\em D80}, 104016.

\end{thebibliography}

\end{document}